# Influence of experimental noise on densities reconstructed from line projections

**M. Samsel-Czekała**[*,1] and **Ł. Boguszewicz**[2]

[1] W. Trzebiatowski Institute of Low Temperature and Structure Research, Polish Academy of Sciences, P.O. Box 1410, 50-950 Wrocław 2, Poland
[2] Faculty of Physics and Astronomy, Wrocław University, sq. M. Born 9, 50-204 Wrocław, Poland

The influence of experimental noise on densities $\rho(\boldsymbol{p})$ reconstructed by the Cormack method from their line projections, e.g. 2D ACAR spectra, is investigated. Simulations of statistical noise are performed for various sets of 2D spectra for two model $\rho(\boldsymbol{p})$ having the cubic symmetry. For the reconstructed densities propagation of the statistical error in terms of standard deviations, $\sigma[\rho(\boldsymbol{p})]$, is estimated. We observe that the distribution of $\sigma[\rho(\boldsymbol{p})]$ has its extremes along the main symmetry directions and also a tendency to accumulate for small $p$. Moreover, the more density components, $\rho_n(p)$, have to be taken to description of $\rho(\boldsymbol{p})$ the less anisotropic is the distribution of $\sigma[\rho(\boldsymbol{p})]$. Additionally, the error generated by the reconstruction method itself is discussed.

**PACS** 05.40.Ca, 06.20.Dk, 71.20.-b, 78.70.Bj, 87.59.Fm

## 1 Introduction

In studies of two-dimensional angular correlation of positron annihilation radiation (2D ACAR) [1] one measures spectra that are line integrals of electron-positron momentum densities $\rho(\boldsymbol{p})$ in the extended zone $\boldsymbol{p}$:

$$J(p_y, p_z) = \int_{-\infty}^{\infty} dp_x \rho(\boldsymbol{p}) . \tag{1}$$

Due to the reconstruction of $\rho(\boldsymbol{p})$ (the same mathematical problem as in computerized tomography [2–4]) information on the electronic structure is obtained. Obviously, the knowledge of the errors contaminating $\rho(\boldsymbol{p})$ is essential for the interpretation of the results. These errors are generated by reconstruction methods themselves as well as follow from experimental statistical noise in $J(p_y,p_z)$.

The problem of the influence of experimental noise on densities reconstructed from projections measured in medical diagnostic studies has been investigated many times, e.g. Ref. [2] and references therein. For $\rho(\boldsymbol{p})$ in solids, where contrary to medical investigations, one uses the symmetry of studied objects, the problem is a little different and has been extensively studied only for $\rho(\boldsymbol{p})$ reconstructed from plane projections (Compton profiles or 1D ACAR spectra) [5–9]

$$J(p_z) = \int_{-\infty}^{\infty}\int_{-\infty}^{\infty} dp_x dp_y \rho(\boldsymbol{p}) . \tag{2}$$

Some questions concerning the influence of noise on $\rho(\boldsymbol{p})$ reconstructed both from plane and line projections are considered in Ref. [10].

---
[*] Corresponding author: e-mail: samsel@int.pan.wroc.pl, Phone: +48 71 343 50 21, Fax: +48 71 344 10 29

In this paper we investigate the propagation of noise for line projections by simulations of statistical noise for various sets of the spectra created for two different model $\rho(p)$ having the cubic symmetry. The densities are reconstructed using the Cormack method [11] and the error distribution in terms of standard deviations, $\sigma[\rho(p)]$, is estimated. For comparison propagation of an error connected with the reconstruction method itself is considered.

## 2 Applied methods

Usually, reconstruction of 3D density $\rho(p)$ resolves itself to separate reconstructions of its 2D subsections from 1D line integrals. Different reconstruction techniques applied in computerized tomography can be employed [2-4,11]. In the Cormack method [11], used by us, both functions $J$ and $\rho$ (in Eq. (1)) are expanded into the Fourier series. By choosing planes $p_z = const.$, perpendicular to the main rotation axis of the crystal, the series reduce to the cosine series [11]

$$\rho(p_x, p_y, p_z = const.) \equiv \rho(p,\varphi) = \sum_{n=0}^{\infty} \rho_n(p)\cos(n\varphi), \qquad (3)$$

$$J(p_y, p_z = const.) \equiv J(t,\beta) = \sum_{n=0}^{\infty} J_n(t)\cos(n\beta), \qquad (4)$$

with $n = iR$ where $R$ is the order of the main rotation axis [001] ($i = 0, 1, 2, \ldots etc.$). $\rho$ and $J$ are described in the polar coordinate systems $(p,\varphi)$ and $(t=|p_y|,\beta)$, respectively, defined on each of the planes $p_z = const.$ Variables $t$ and $\beta$ denote, respectively, the distance of the integration line from the origin of the polar system and its angle with respect to the fixed axis $p_x$, while $p = \sqrt{p_x^2 + p_y^2}$ is a radial momentum on each plane.

In 1964 Cormack showed that if $J_n(t)$ is expanded into a series of Chebyshev polynomials of the second kind [$U_l(t)$]

$$J_n(t) = 2\sum_{m}^{\infty} a_n^m \sqrt{1-t^2} U_{n+2m}(t), \qquad (5)$$

Eq. (1) can be solved analytically and

$$\rho_n(p) = \sum_{m}^{\infty} (n+2m+1) a_n^m R_n^m(p), \qquad (6)$$

where $R_n^m(p)$ are Zernike polynomials and $a_n^m$ are evaluated from Eq. (5) utilizing the orthogonality relation for $U_l(t)$.

In order to study propagation of experimental noise in densities $\rho(p)$ reconstructed from line projections, we created two kinds of model $\rho(p)$ having the cubic symmetry, presented in Fig. 1. 1D spectra of each model density, $\rho_{mod}(p)$, were calculated for different angles $\beta$, equally spaced ($\Delta\beta = const.$) in the range of the irreducible part of the plane (from $\beta = 0°$ to $45°$ in the case of the cubic or tetragonal structures). For each $\rho_{mod}(p)$ we created a series of 5 spectra ($\Delta\beta = 11.25°$) and, additionally for the model II, a set of 10 spectra ($\Delta\beta = 5°$). For each projection we performed $M = 20$ different simulations of statistical noise, employing a gaussian random number generator with standard deviation $\sigma = \sqrt{N}$ where $N$ denotes the total number of counts per sampling point.

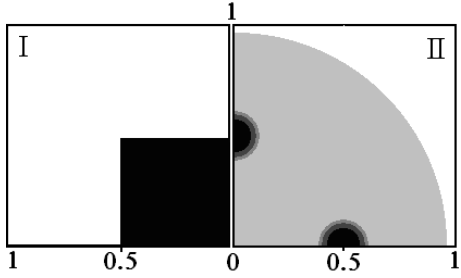

**Fig. 1** Two model cubic densities $\rho(p)$ (I and II) in the (001) plane. The darker colours denote the higher values. Black areas correspond to: 100000 and 50000 for model I and II, respectively, while white areas denote 0. Densities (for each model) are displayed in the range (1.0 ×1.0).

In this way, for each series of model spectra, we created 20 sets of the spectra with different noise distributions. Then, 20 sets of densities $\rho_i(p)$ (in the (001) plane) were reconstructed (in the circle with the radius $p = 1$) using the Cormack method [11]. Finally, we evaluated the error propagation in terms of standard deviations $\sigma = \sigma[\rho(p)]$ using estimators [9]

$$\sigma[\rho(p)] = \sqrt{(1/M)\sum_{i=1}^{M}[\rho_i(p) - \overline{\rho}(p)]^2} \qquad (7)$$

where the average value of the densities with noise, $\overline{\rho}(p)$, should be equal to $\rho_{mod}(p)$. Generally, if $\overline{\rho}(p)$ are unknown, they can be approximated via estimators $\overline{\rho}(p) = (1/M)\sum_{i=1}^{M}\rho_i(p)$. To investigate how far the noise is filtered by the expansion into orthogonal (this case Chebyshev) polynomials [Eq. (5)], the error distributions $\sigma[\rho(p)]$ were calculated for different numbers of coefficients $a_n^m$ (20, 30, 45, 60, 90, 150, 300). We evaluated also the contribution of the error connected with the isotropic and anisotropic part of the densities, i.e. $\sigma[\rho_0(p)]$ and $\sigma[\rho_a(p)]$ where $\rho_a(p) = \rho(p) - \rho_0(p)$, as well as connected with particular density components, $\sigma[\rho_n(p)\cos(n\varphi)]$ – see Eq. (3).

Furthermore, the error connected with the reconstruction method itself was determined as $Err_{rec}[\rho(p)] = |\rho_{mod}(p) - \rho_{rec}(p)|$ where $\rho_{rec}(p)$ are densities reconstructed from the model spectra without noise.

## 3  Results

In the case of the model I, the errors $\sigma[\rho(p)]$ and $\sigma[\rho_a(p)]$ are presented in Fig. 2 for chosen numbers of coefficients $a_n^m$, i.e. 45 and 60. In the central square (region where model $\rho(p) \neq 0$) the error distributions have pronounced maxima along the high symmetry [100] and [110] directions − this behaviour does not depend on the number of $a_n^m$ used to description of $\rho(p)$ and is connected with the following.

Each spectrum contains noise that, of course, has no symmetry. However, projections are measured in the irreducible part of space (this case of (001) plane), so the spectra (thus also the noise) as well as the reconstructed densities have to be expanded into the lattice harmonics (this case cosine Eqs. (3) and (4)) series. Due to this the symmetry is imposed both on the data and noise, as noticed in Ref. [9]. In the case of plane projections we have to describe $\rho(p)$ by the lattice harmonics, which have their extremes for high symmetry directions in 3D space (in particular [001] and [111] for the cubic structures). In this way the symmetry is imposed also on the noise propagated with radial density components and, as a result, the noise distribution $\sigma[\rho(\mathbf{p})]$ should have maxima along the same high symmetry directions [5–9]. In the case of line projections, if the reconstruction is performed separately on each of independent planes, the expansion of $\rho(p)$ reduces to the cosine series (Eq. [3]) which terms have extremes for $|\cos(n\varphi)| = 1$ that is satisfied for $\varphi$ equally spaced from 0 at $\Delta\varphi = \pi/n$. Thus, the noise distribution connected with each density component $\sigma[\rho_n(p)\cos(n\varphi)]$ has maxima along the same directions as $|\cos(n\varphi)|$, which is shown in Fig. 3 on the examples of $n = 4$ and 8. Since all terms $|\cos(n\varphi)|$ have their maxima also for the main symmetry directions, i.e. for $\varphi = 0$ and $\pi/R$ (where $R = 3$, 4, and 6 for the trigonal, tetragonal or cubic, and hexagonal symmetries, respectively), the total noise distribution $\sigma[\rho(\boldsymbol{p})]$ ought to have maxima just along these directions.

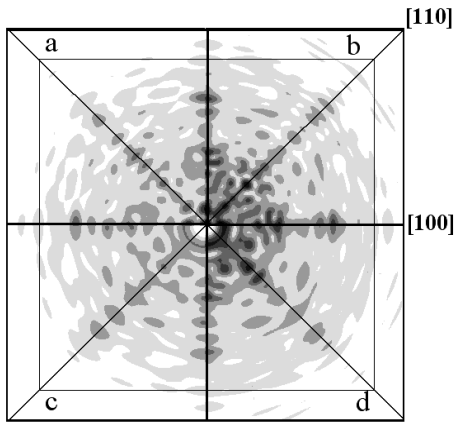

**Fig.** Error distributions $\sigma[\rho(p)]$ (upper quadrants) and $\sigma[\rho_a(p)]$ (bottom quadrants) for $\rho(p)$ reconstructed from the sets of 5 spectra of the model I, using 45 (a,c) and 60 (b,d) coefficients $a_n^m$. The darker colours denote the higher values. Results (in each quadrant) are displayed in the range (0.6 × 0.6).

For more detailed analysis of the results, in Fig. 4 we present $\sigma$, the same as in Fig. 2, but only along the main symmetry directions, i.e. [100] and [110]. Additionally, the part of noise propagated with the isotropic component, $\sigma[\rho_0(p)]$, is displayed. These figures indicate that the smaller number of $a_n^m$ the lower values of $\sigma$. It can be explained by the fact that the expansion of measured spectra into orthogonal (Chebyshev) polynomials [Eq. (5)] has mean-squares approximation properties. Thus, when employed a limited number of its coefficients, $a_n^m$, the experimental noise in the data is effectively smoothed [12].

Of course, in practise an optimum number of $a_n^m$ cannot be too small in order not to smear essential details of the reconstructed densities (see Fig. 7). Moreover, some part of noise is eliminated via the reconstruction procedure which naturally imposes the requirement of the consistency of spectra that represent integrals of the same density [13].

It is also seen that for these model densities, having a relatively strong anisotropy, $\sigma[\rho_a(\boldsymbol{p})]$ has comparable values to $\sigma[\rho(\boldsymbol{p})]$, except for the range of small $p$, where $\sigma[\rho_0(p)]$ has extremely high values.

In the case of the model II we present, in Figs. 5 and 6, $\sigma[\rho(\boldsymbol{p})]$ for $\rho(\boldsymbol{p})$ described by 5 and 10 density components $\rho_n(p)$, reconstructed from the set of 5 and 10 spectra, respectively, using 30 and 60 coefficients $a_n^m$. It is seen that also for this model the smaller number of $a_n^m$ the lower values of $\sigma$ as well as σ[ρ(***p***)] is the highest along the high symmetry directions ([100] and [110]). Furthermore, in the case of 10 spectra σ[ρ(***p***)] is less anisotropic than that obtained for 5 spectra.

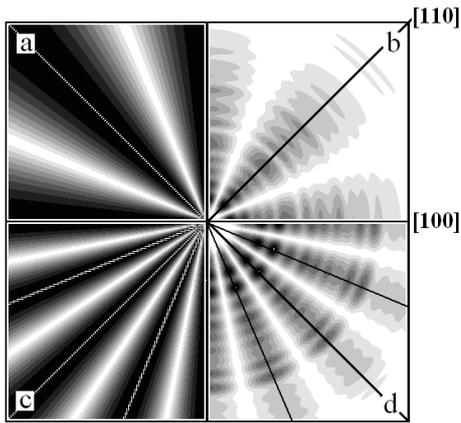

**Fig. 3** Terms $|\cos(n\varphi)|$ for $n=4$ (a) and 8 (c) and error distributions $\sigma[\rho_n(p)\cos(n\varphi)]$ for $n=4$ (b) and 8 (d). Functions in all quadrants are drawn separately (in different scales of colours). Their maximum values are marked by radial lines and the darker colours denote the higher values. $\sigma$ is displayed in the same range as in Fig. 2.

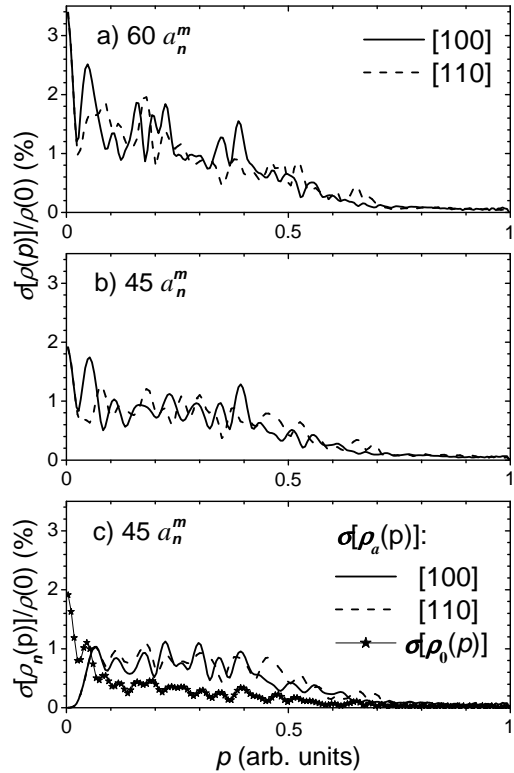

**Fig. 4** The same $\sigma$ as in Fig. 2 but only along the main symmetry directions and in % of $\rho_0(0)$. Additionally, $\sigma[\rho_0(p)]$, is displayed.

The reason for such a behaviour is the following. $\rho_n(p)$ and thus also the noise are multiplied by terms $\cos(n\varphi)$ having extremes for $\varphi$ spaced at $\Delta\varphi = \pi/n$ − the higher $n$ the more directions $\varphi$ with the extremes (compare cases (a) and (c) in Fig. 3). As a result, noise contributions connected with $\rho_n(p)\cos(n\varphi)$ for higher $n$ are less anisotropic. Therefore, if a large number of components $\rho_n(p)$ have to be used to description of ρ(***p***), the total error distribution, σ[ρ(***p***)], becomes also relatively less anisotropic.

Moreover, in the case of 10 spectra σ[ρ(***p***)] has much lower values than σ[ρ(***p***)] obtained from 5 spectra – fluctuations seen in bottom quadrants of Fig. 5 would be invisible when drawn in the same scale as for the upper quadrants – the scales in Fig. 6 also differ from each other. This behaviour is simple to explain [10]. Namely, by utilizing more projections (measured with the same statistics) one gets better total statistics of measurements and thus σ[ρ(***p***)] is lower.

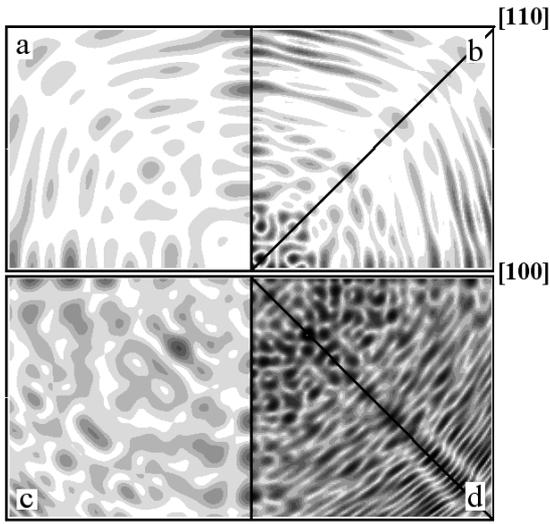

**Fig. 5** Error distributions σ[ρ(***p***)] for ρ(***p***) reconstructed from the set of 5 (upper quadrants) and 10 (bottom quadrants) spectra of the model II, using 30 (a,c) and 60 (b,d) $a_n^m$. Results in cases (a,b) are drawn separately (different scales of colours) from those for (c,d) and are displayed (in each quadrant) in the range $(0.675 \times 0.675)$.

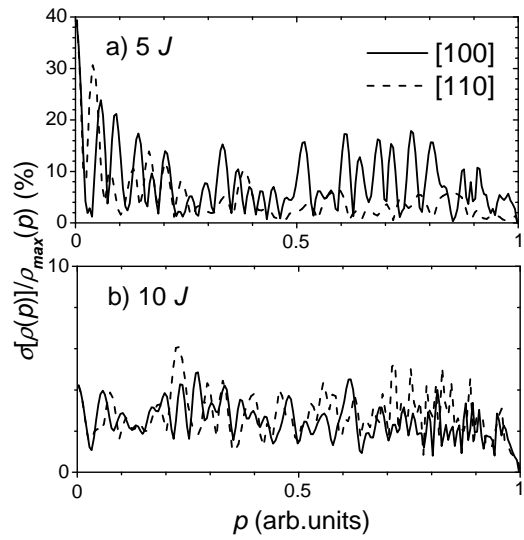

**Fig. 6** The same as in Fig. 5 but only for the high symmetry directions and the case of 60 $a_n^m$, in % of $\rho_{\max}(p)$. σ[ρ(***p***)] for ρ(***p***) reconstructed from the sets of 5 and 10 spectra are displayed in parts (a) and (b), respectively, in different scales.

For comparison, in Figs. 7 and 8 the error generated by the reconstruction method itself, $Err_{rec}[\rho(\mathbf{p})]$, for Model II is displayed. This error does not exhibit the highest oscillations along the main symmetry directions. The distribution of $Err_{rec}[\rho(\mathbf{p})]$ is strongly dependent (contrary to σ[ρ(***p***)]) on the particular shape of the model, having in this case maxima around the "sharp" edges of spheres, as well as on the number of projection used to reconstruction.

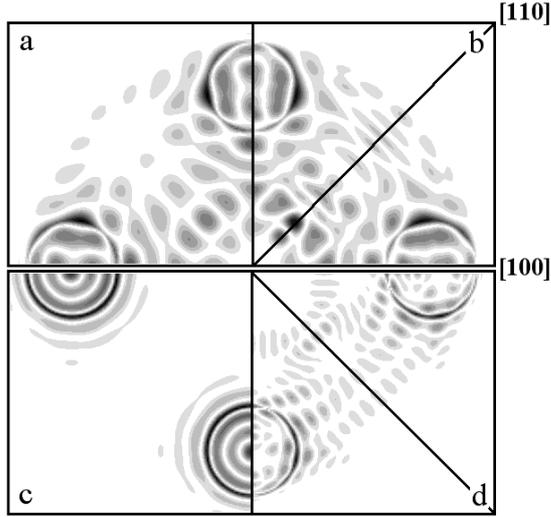 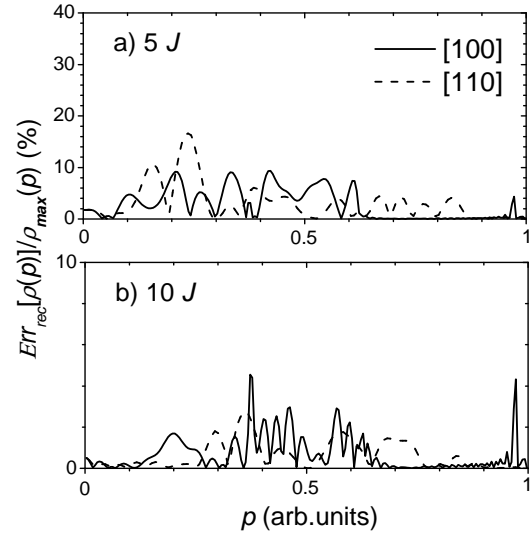

**Fig. 7** The same as in Fig. 5 but for the reconstruction error $Err_{rec}[\rho(\boldsymbol{p})]$.

**Fig. 8** The same as in Fig. 6 but for $Err_{rec}[\rho(\boldsymbol{p})]$.

## 4 Conclusions

We proved that the distribution of $\sigma[\rho(\boldsymbol{p})]$ for $\rho(\boldsymbol{p})$ reconstructed from line projections has its extremes along the high symmetry directions (opposed to $Err_{rec}[\rho(\boldsymbol{p})]$) and also a tendency to accumulate for small $p$, in agreement with the results for plane projections [5–9]. Moreover, we observed that the more density components $\rho_n(p)$ is taken to description of $\rho(\boldsymbol{p})$ the less anisotropic is the distribution of $\sigma[\rho(\boldsymbol{p})]$. Furthermore, $\sigma[\rho(\boldsymbol{p})]$ can be minimized, if one uses such reconstruction techniques in which measured spectra are expanded into orthogonal polynomials.

**Acknowledgements** We are very grateful to Prof. G. Kontrym-Sznajd for giving us the inspiration to write this paper and for critical comments which led to its improvement as well as to the Polish State Committee for Scientific Research (Grant No. 2 P03B 012 25) for financial support.